\documentstyle[11pt,newpasp,twoside,psfig]{article} 
\def\edcomment#1{\iffalse\marginpar{\raggedright\sl#1\/}\else\relax\fi} 
\reversemarginpar 

\begin{document} 
\title{Observational Properties of Extragalactic Globular
           Cluster Systems}

\author{Duncan A. Forbes} 
\affil{Centre for 
Astrophysics \& Supercomputing, Swinburne University, Hawthorn VIC 
3122, Australia} 

\begin{abstract} 
The superior resolution of HST and the light gathering power of 
large 8-10m class telescopes are now providing information on
distant globular clusters (GCs) that is comparable to that
obtained in early 1990s for Local Group systems. Here I summarise
what has been learnt from the imaging and limited spectroscopy of
GCs in other galaxies. The GC systems of spirals and ellipticals
reveal remarkable similarities. The vast bulk of GCs appear to
have formed at early epochs, with mergers making a
limited contribution to the overall GC system at later epochs. 
These observational findings 
are placed in the context of galaxy formation. 

\end{abstract}

\section{Introduction} 

Much has been learnt about the formation and evolution of our
Milky Way Galaxy from the system of globular clusters (GCs). It
is natural to extend such studies to the GC systems of other
galaxies. There are several advantages in doing this:\\ $\bullet$
Greater numbers, eg the MW has only 150 known GCs, M31 has a
total population of $\sim$
400 and M87 $\sim$ 10,000.\\ 
$\bullet$ Probing to higher metallicity GCs.\\
$\bullet$ The Local Group has no
giant ellipticals.\\ 
$\bullet$ No foreground stars in the GC spectra.\\ $\bullet$
Discover something new !\\

For Local Group studies (D $\sim$ 1 Mpc), 3-6m telescopes have
collected optical and IR colours, and measured sizes for large
numbers of GCs. Spectra also exist in large numbers for V $\sim$
16 GCs, giving metallicities, abundances, relative ages and
system kinematics. Classic papers for Local Group GCs are those
of Brodie \& Huchra in 1990s (eg Brodie \& Huchra 1990, 1991)
More recently the 0.1$^{''}$ resolution of HST has provided CMD
morphologies for several Local Group GCs (see various posters at
this conference).

Further afield (eg Virgo/Fornax distance of 15 Mpc) 8-10m
telescopes are starting to obtain similar information for more
distant GC systems. Optical colours have now been measured for $\sim$50
systems. IR colours are lagging behind but quickly catching up
(see posters by Puzia and Hempel). At 15 Mpc, GCs are partially
resolved by HST allowing sizes to be measured (eg Larsen et
al. 2001). Typical mags of V $\sim$ 22 have made obtaining high
quality spectra (for ages, abundances etc) and large numbers for
system kinematics slow going.  Only a handful of systems have
good quality spectra for individual GCs (the vast majority coming
from the SAGES project.) A database of imaging and spectral
information for extragalactic GCs can be found at:
astronomy.swin.edu.au/dforbes

\section{Bulge GCs in Spirals}

In our Galaxy, the inner metal-rich GCs are now thought to be
associated with the bulge rather than the disk (Minniti
1995). Recently Forbes et al. (2001) extended this view to other
spiral galaxies. In particular, they showed that the mean
metallicity, velocity dispersion, rotation velocity and spatial
distribution of the inner metal-rich GCs in M31 and M81 supported
a bulge origin. HST imaging of the GCs in the Sombrero galaxy
(M104) revealed a large number of metal-rich GCs (Larsen, Forbes
\& Brodie 2001). It seems highly unlikely that they are associated with
the tiny disk but instead with the dominant bulge. A comparison
of the number of metal-rich GCs with the bulge luminosity (called
`bulge' specific frequency) in the MW, M31 and M104 suggests a
near constant value of 1.0. This can be compared to the specific
frequency of red (metal-rich) GCs in field ellipticals which is
0.5--1.5 (assuming that the elliptical is simply a large
bulge). Thus the `bulge specific frequency' for spirals and
ellipticals are similar.

\section{The Local Group Elliptical}

Some, perhaps all, large ellipticals are thought to have formed
via the merger of spirals. It is thus interesting to `merge'
together the GC systems of the Local Group galaxies to simulate a
dissipationless (ie no new GCs are formed) merger. This merger is
of course dominated by M31 (Sb) and MW (Sbc).  Forbes et
al. (2000) carried out this exercise and found the total GC
system of the LG elliptical to be 700 $\pm$ 125. The total
luminosity was estimated to be M$_V$ --22.0, giving S$_N$ =
1.1. After correcting for the presence of young stellar
populations this value rises to S$_N$ $\sim$ 2, so still a
relatively low S$_N$ elliptical. The GC luminosity function
revealed a peak at the universal value (ie M$_V$ $\sim$ --7.5)
and the metallicity distribution was bimodal with peaks [Fe/H]
$\sim$ --1.5 and --0.5. The ratio was 2.5:1 for metal-poor to
metal-rich. Any models of merging spirals should take into
account not only the pre-existing metal-poor GCs but also the
metal-rich ones (which could be a significant fraction in the
case of early-type spiral progenitors).

\section{Overview of GC imaging}

\noindent
$\bullet$ Metallicity distributions. All galaxies, irrespective
of Hubble type, appear to have a population of metal-poor ([Fe/H]
$\sim$ --1.5) GCs. Galaxies with bulges have a metal-rich ([Fe/H]
$\sim$ --0.5) population also. Some particular exceptions may
exist (eg Woodworth \& Harris 2000).  
.\\ $\bullet$ Metallicity-mass relation. The mean
metallicity of the metal-poor GCs is almost constant at [Fe/H]
$\sim$ --1.5, whereas the mean value for the metal-rich ones
varies with host galaxy mass (ie luminosity or velocity
dispersion). This suggests that the metal-poor ones formed
slightly pre-galactic while the metal-rich ones know about the
potential that formed they in (Forbes \& Forte 2001).\\ $\bullet$
Size trend. Metal-poor GCs are, on average, 20\% larger than
metal-rich ones irrespective of Hubble type (Larsen et
al. 2001). This effect may be due to conditions at formation or
subsequent dynamical evolution due to different orbits. \\
$\bullet$ Spatial distribution. In general, the metal-rich GCs in
ellipticals are centrally concentrated and appear to follow the
`bulge/spheroid' light. Whereas the metal-poor GCs are more
extended and are associated with the halo.\\

\section{Overview of GC Spectroscopy}

\noindent
$\bullet$ Metallicities. Spectra have confirmed the bimodality
seen in optical colours and the metallicity range of less than
one hundredth solar to twice solar. \\ $\bullet$ Ages. Relative
ages, from Lick indices, are now available for a number of GCs in
ellipticals. These suggest that most GCs are old ($\sim$ 12 Gyrs)
although there is tentative evidence that the red GCs may be a
few Gyrs younger than the blue ones. There are also cases now of
some intermediate (Puzia et al. 2002; Larsen
et al. 2002) and young age GCs in otherwise `old' ellipticals 
(Forbes et al. 2001). It is not yet clear how significant this
younger 
subpopulation is compared to the total. One caveat is the possible
presence of blue horizontal branches which may affect the age
determinations based on the H$\beta$ line (Lee et al. 2000).\\ $\bullet$
Abundances. To date most GCs in ellipticals are consistent with
supersolar abundances, ie [Mg/Fe] $\sim$ +0.3 (although 
Maraston et al. 2001 found solar ratios for GCs in the ongoing
merger NGC 7252). This indicates the dominance of SN II over SN
Ia in their formation, which is probably due to a short time
scale for star formation. Further abundance studies will reveal
important chemical evolution clues. \\

\noindent
Currently all high quality spectra (ie H$\beta$ errors of $< \pm
0.3\AA$) of GCs beyond the Local Group come from the Keck
telescope. These include:\\
NGC~1399 Forbes et al. (2001)\\
M81 Schroder et al. (2002)\\
M104 Larsen et al. (2002)\\
NGC~524 Beasley et al. (2003)\\
NGC~4365 Larsen et al. (2002)\\
In the near future, spectra from the VLT and Gemini telescopes
will add to this list. 

\begin{figure}[!t]
\centerline{\vbox{}\psfig{figure=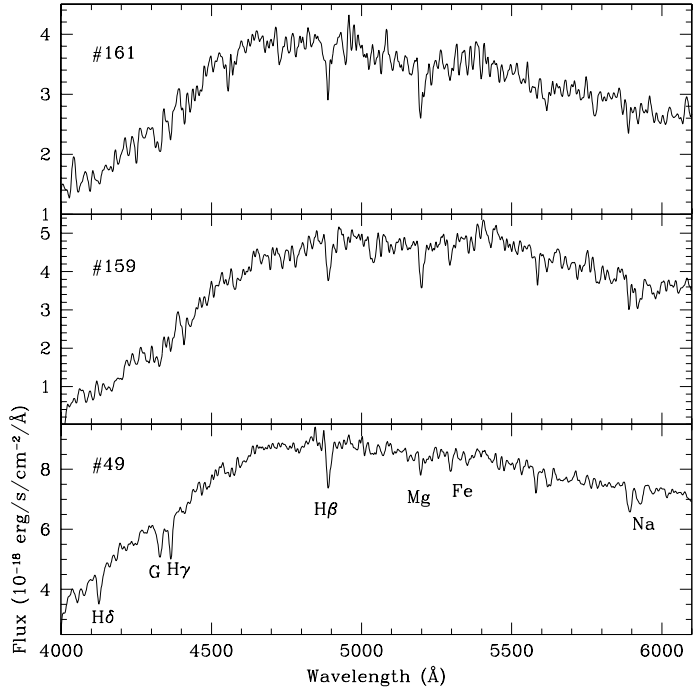,height=10cm,width=10cm,angle=0}}
  \caption{Examples of Keck spectra for globular clusters in NGC~1399. 
These spectra represent 3.5 hours of Keck/LRIS time, and have
H$\beta$ 
errors 
of $\sim$0.25~\AA~ (about $\pm$4 Gyrs in age). 
}
  \label{Figure1}
\end{figure}

An example of three GCs in NGC~1399 observed by Keck + LRIS is
shown in Fig. 1. Here the exposure time was 3.5 hours and the
typical H$\beta$ error was $\pm$0.2--0.4\AA. For NGC~1399 we
found (Forbes et al. 2001) that the GCs covered the full
metallicity range, ie --2.2 $<$ [Fe/H] $<$ 0.3, and that most of
the blue and red GCs were old ($\sim$ 12 Gyrs). However we also found
two young ($\sim$ 2 Gyrs) GCs. All of the GCs were consistent
with having supersolar [Mg/Fe] abundance ratios.

GC system kinematics have only been measured for a small number
of elliptical galaxies (eg Kissler-Patig \& Gebhardt 1988; Zepf et al. 2000), and each
system appears to have slightly different kinematic trends.  So
it is difficult to draw general conclusions. This situation
should change rapidly with new spectrographs coming online which
have hundreds of slits (eg DEIMOS, GMOS, VIMOS). As well as
constraining galaxy formation models, it will be interesting to
compare mass estimates from GCs (which require assumptions about
orbits) with those from X-ray measurements (which requires the
assumption of hydrostatic equilibrium).

\section{The Formation and Evolution of GCs}

Three scenarios had been proposed to understand the observations
of extragalactic GC systems in the context of galaxy
formation. They are late stage mergers (Ashman \& Zepf 1992), two-phase
collapse at early epochs (Forbes, Brodie \& Grillmair 1997) and accretion (Cote,
Marzke \& West 1998).  Galaxy formation itself is often
interpreted using a semi-analytic code (eg Cole et
al. 2000) within a $\Lambda$CDM universe. Recently we simulated the
formation of GCs around ellipticals using the GALFORM
semi-analytic code (eg Cole et al. 2000) in such a universe for
the first time. This model involves elements of mergers, collapse
and accretion.

\begin{figure*}
\centerline{\vbox{}\psfig{figure=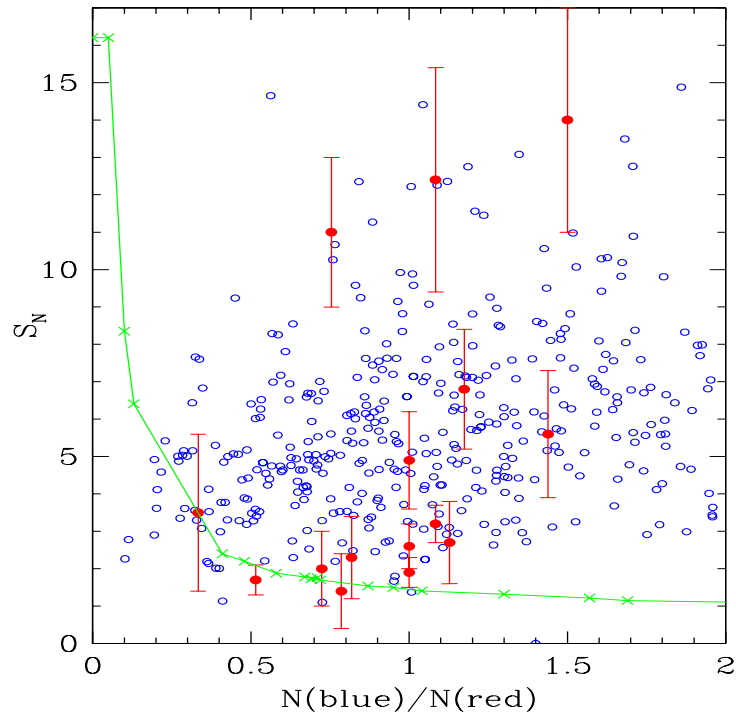,height=10cm,width=10cm,angle=0}}
\caption{Specific frequency vs ratio of blue to red globular clusters. 
The small open circles represent the model galaxies from GALFORM,
the filled circles with error bars are observed galaxies for
which the imaging covers a wide area (so this excludes most HST
studies), and the solid line shows the spiral-spiral merger
simulations of Bekki et al. (2002) for various impact
parameters. As suggested by the merger simulations, the
expectation is that high S$_N$ galaxies will have low blue-to-red
ratios. This is contrary to the general trend seen in the data
and in the GALFORM model galaxies. }
  \label{Figure2}
\end{figure*}

Details of this work can be found in Beasley et
al. (2002). Briefly, GCs are formed in two modes of star
formation. In the first, or `quiescent' mode, metal-poor GCs form
in proto-galactic clouds. These gaseous clouds collapse/merge,
giving rise to a burst of star formation. During this `burst'
mode, the vast bulk of the galaxy stars form along with the
metal-rich GCs. The final GC system depends on the local galaxy
environment and its mass (luminosity). For example, low
luminosity field ellipticals tend to have a more extended star
formation history which is more bursty in nature. We would expect
such galaxies to reveal metal-rich GCs that are 3-5 Gyrs younger
in the mean than the metal-poor ones which are $\sim$12 Gyrs old
in all ellipticals. More high quality spectra with 8-10m class
telescopes should be able to test this prediction.

We had to make two key assumptions in order to produce bimodal GC
colour distributions.  The first was to fix the efficiency of
forming GCs vs field stars (this was chosen to match observations
of M49). The second was to stop the formation of metal-poor GCs
at redshift z = 5. There was no physical basis for this, but such
a redshift may be associated with the epoch of reionisation (see
Cen 2001). The model ellipticals have GC systems that can
reproduce the diversity seen in HST studies (eg Larsen et
al. 2001) and the relation between number of GCs and galaxy
luminosity (ie typical S$_N$ = 5).

In Fig. 2 we plot S$_N$ (number of GCs per unit galaxy starlight)
vs the ratio of blue (metal-poor) to red
(metal-rich) GCs. We compare the model predictions of 450
ellipticals with galaxies for which wide field imaging data is
available. We also show the range of parameters for the major
merger model of Bekki et al. (2002). The GALFORM model and
the data both show a weak trend for the higher S$_N$ galaxies to
have relatively more blue GCs, ie it is the number of blue GCs
that drives S$_N$. Examination of the evolution of S$_N$ with
time in individual galaxies also indicates that most of the
change in S$_N$ occurs at early epochs and late stage mergers
have only a minor effect on S$_N$ (see Fig. 3). 

\begin{figure}[!t]
\centerline{\vbox{}\psfig{figure=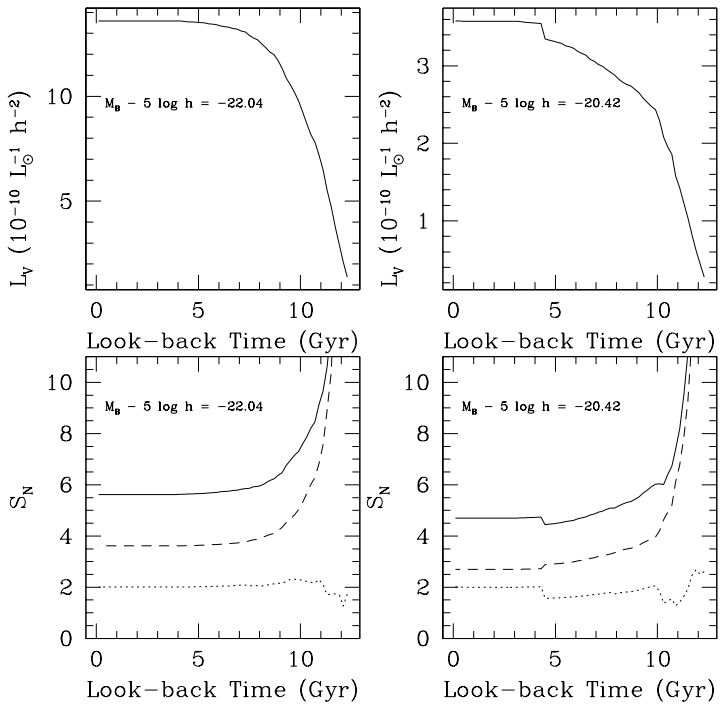,height=10cm,width=10cm,angle=0}}
  \caption{Evolution of galaxy luminosity and 
globular cluster specific frequency S$_N$ with
look back time for two different galaxy models from
GALFORM. The dotted line shows the evolution of S$_N$ for the red
clusters, the dashed line for blue clusters, and the solid line
for the total cluster system. The lower luminosity galaxy undergoes a
merger at $\sim$5 Gyrs ago, which causes only a small increase in
the overall S$_N$. 
}
  \label{Figure3}
\end{figure}

On the other hand, major mergers would require a different
trend. To form a high S$_N$ system via mergers one needs many red
GCs. The major merger predictions of Bekki et al. (2002) are
clearly at odds with the data, suggesting that such mergers are
not the dominant process in creating high S$_N$ galaxies. Indeed
observations of young (few Gyr old) ellipticals suggest S$_N$
values close to 2--3 (after correction for the M/L ratio of the
younger stellar populations).

\section{Summary and Predictions}

\noindent
The observations of extragalactic GC systems indicate that:\\
$\bullet$ The inner metal-rich GCs in the Milky Way and other
spirals has a bulge (not disk) origin.\\ $\bullet$ The GC systems
of spirals and ellipticals show remarkable similarities.\\
$\bullet$ Blue and red GCs have similar ages $\sim$ 12 Gyrs (but
red GCs could be younger by 2--4 Gyrs).\\ $\bullet$ Some young
($\sim$ 2-8 Gyrs) GCs have been found in `old' ellipticals.\\
$\bullet$ Blue and red GCs appear to have supersolar alpha ratios.\\
$\bullet$ Red GCs trace elliptical galaxy star formation.\\

\noindent
The modelling of GC formation in a $\Lambda$CDM
universe indicates that:\\ $\bullet$ Blue
GCs formed $\sim$ 12 Gyrs ago in all ellipticals.\\ $\bullet$ Red
GCs have a mean age of 8--10 Gyrs in field 
ellipticals.\\ $\bullet$ S$_N$ is driven by the number of blue
GCs in a galaxy and is largely determined at early epochs; late
stage mergers have little effect on S$_N$.\\ $\bullet$ In
general, spirals and ellipticals have similar GC systems.\\

\noindent
Areas of research that we will no doubt hear
more about in the near future are:\\ $\bullet$ 
8m wide field K band imaging studies (eg Puzia et al. 2002).\\
$\bullet$ Age structure within the red GCs (Beasley et
al. 2002).\\ $\bullet$ Nature of the large (R$_{eff}$ $\sim$ 10
pc) red low luminosity (M$_V$ $\sim$ --6) clusters (Brodie this
meeting).\\ $\bullet$ Importance of shredded dwarf galaxies (eg
Bekki et al. 2001).\\ $\bullet$ HST+ACS CMDs for Local Group GCs
(eg Rich this meeting).\\ $\bullet$ Halo mass estimates, GC
kinematics vs Xrays (eg Kissler-Patig et al. 1998).\\ $\bullet$
Continued improvements in SSP grids (eg Maraston this meeting)\\

\section{References}

\noindent
Ashman, K., Zepf, S., 1992, ApJ, 384, 50\\ 
Beasley, M., Baugh,
C., Forbes, D., Sharples, R., Frenk, C., 2002, MNRAS, 333, 383\\
Beasley, M., et al. 2003, MNRAS, submitted\\ 
Bekki, K., Couch, W.,
Drinkwater, M., 2001, ApJL, 552, 105\\ Bekki, K., Forbes, D.,
Beasley, M., Couch, W., 2002, MNRAS, submitted\\ 
Brodie, J., Huchra, J., 1990, ApJ, 362, 503\\
Brodie, J., Huchra, J., 1990, ApJ, 379, 157\\
Cen, R., 2001, ApJ, 560, 592\\
Cole, S., Lacey, C., Baugh, C., Frenk, C., 2000, MNRAS, 319, 168\\ 
Cote, P., 
Marzke, R., \& West, M., 1998, ApJ, 501, 554\\ 
Forbes, D., Brodie, J., Grillmair, C., 1997, AJ, 113, 1652\\ 
Forbes, D.,
Brodie, J., Larsen, S., 2001, ApJL, 556, 83\\ 
Forbes, D., Forte,
J., 2001, MNRAS, 322, 257\\ Forbes, D., Beasley, M., Brodie, J.,
Kissler-Patig, M., 2001, ApJL, 563, 143\\ 
Kissler-Patig, M., Gebhardt, K., 1998, AJ, 116, 2237\\
Larsen, S., Brodie, J., Huchra, J., Forbes,
D., Grillmair, C., 2001, AJ, 121, 2974\\
Kissler-Patig, M.,
Brodie, J., Schroder, L., Forbes, D., Grillmair, C., Huchra, J.,
1998, AJ, 115, 105\\ 
Larsen, S., Forbes, D., Brodie, J., 2001, MNRAS, 327, 1116\\ 
Larsen, S., et al. 2002, MNRAS, in press\\ 
Lee, H.C., Yoon, S.J., Lee, Y.W., 2000, AJ, 120, 998\\
Maraston, C., et al. 2001, A\&A, 370, 176\\
Minniti, D.,
1995, AJ, 109, 1663\\ 
Puzia, T., Zepf, S., Kissler-Patig, M.,
Hilker, M., Minniti, D., Goudfrooij, P., 2002, astro-ph/0206147\\
Schroder, L., et al. 2002, AJ, 123, 2473\\
Woodworth, S., Harris, W., 2000, AJ, 119, 2699\\
Zepf, S., et al. 2000, AJ, 120, 2928\\

\end{document}